\documentclass[manuscript]{aastex}

\newcommand{\nmem}{$N_{mem}$~}
\newcommand{\kms}{km s$^{-1}$~}
\newcommand{\sv}{$\sigma_{cz}~$}
\newcommand{\mss}{$m_{SSRS2}~$}
\newcommand{\msst}{$m_{SSRS2,new}~$}
\newcommand{\mzw}{$m_{Zw}~$}
\newcommand{\uonly}{U$_{only}$~}
\newcommand{\uboth}{U$_{both}$~}

\shorttitle{USGC}
\shortauthors{Ramella et al.}

\slugcomment{Revised February 14, 2002}
\begin{document}

\title{The UZC-SSRS2 Group Catalog}
\author{Massimo Ramella}
\affil{INAF, Osservatorio Astronomico di Trieste}
\affil{via G. B. Tiepolo 11, I-34131 Trieste, Italy}
\email{ramella@ts.astro.it}
\author{Margaret J. Geller}
\affil{Smithsonian Astrophysical Observatory}
\affil{60, Garden Str., Cambridge, MA02138}
\email{mjg@cfa.harvard.edu}
\author{Armando Pisani}
\affil{Istituto di Istruzione Statale Classico Dante Alighieri, 
Scientifico Duca degli Abruzzi, Magistrale S. Slataper}
\affil{viale XX settembre 11, I-34170 Gorizia, Italia}
\email{arpisan@tin.it}
\and
\author{Luiz N. da Costa}
\affil{European Southern Observatory}
\affil{Karl Schwarzschild Str. 2, D-85748 Garching, Germany}
\email{ldacosta@eso.org}
\pagebreak

\begin{abstract}
We apply a friends-of-friends algorithm to the combined UZC and SSRS2 
redshift surveys to construct a catalog of 1168 groups of galaxies; 411 
of these groups have 5 or more members within the redshift survey. The 
group catalog covers 4.69
steradians and all groups exceed the number density contrast threshold, 
$\delta
\rho/\rho = 80$. We demonstrate that the groups catalog is homogeneous 
across
the two underlying redshift surveys; the catalog of groups and their 
members
thus provides a basis for other statistical studies of the large-scale
distribution of groups and their physical properties.  The median 
physical
properties of the groups are similar to those for groups derived from
independent surveys including the ESO Key Programme and the Las Campanas
Redshift Survey. We include tables of groups and their members. 
\end{abstract} 

\section{Introduction}

Catalogs of loose groups have long been a fundamental resource for the 
study of
these abundant systems. Early catalogs constructed with a variety of 
selection
criteria were based on positions of galaxies on the sky \citep{deV74, 
TG76,
Mat78}.  There has been a steady evolution toward groups selected from
redshift surveys of the local universe \citep{HG82, Ven84, NW87, Tul87, 
Mai89,
Ram89, Gou92, Gar93, Ram97, Tra98, Ram99, Giu00, Tuc00, Car01}, with the
notable exception of ~ \citet{Whi99} who use the \citet{TG76} algorithm 
to
select groups projected on the sky starting from the CGCG \citep{Zwi}.

Here we discuss the group catalogs for the UZC \citep{Fal99}
and SSRS2 \citep{daC98} redshift surveys which
cover 4.69 steradians of the sky to a limiting m$_{B(0)} \simeq 15.5$.  
To
identify the groups we apply a standard friends-of-friends algorithm
\citep{HG82}.  Analyses based on preliminary versions of this catalog 
are
already in the literature. ~ \citet{Gir00} and \citet{Pad01} compute 
the group
correlation function.  ~ \citet{Tra97} identify groups in the 
Perseus-Pisces
region and computes the  group correlation function \citep{Tra98}.
\citet{Mah00} cross-correlate a portion of the catalog with the Rosat 
All-Sky
Survey to identify systems which have associated extended x-ray 
emission.
\citet{Fr95a, Fr95b} and \citet{Dia99} compare the properties of these 
systems
with the predictions of n-body simulations. Issues including 
spectroscopic and
morphological types of galaxies in nearby groups remain to be 
investigated.

There is wide variation in the group selection algorithms.  This 
variation
affects the fraction of galaxies assigned to groups. It also affects the
derived physical parameters ~ \citep{Pis92}. \citet{Fr95a, Fr95b}  
discussed the sensitivity of
group identification to the algorithm. A similar discussion is beyond 
the scope
of this paper: all of the redshift survey data are publicly available 
for
examination of these issues.

Here we provide lists of groups and group members to enable reproduction
and extension
of published results along with comparisons to other group catalogs. We 
include
cross-references to our previous catalogs constructed from partial 
versions of
the UZC and SSRS2.  We refer to the entire group catalog discussed in 
this
paper as the {\it UZC-SSRS2 Group Catalog}  which we
abbreviate as USGC. 
Section 2 reviews the redshift surveys and compares the UZC with the 
SSRS2 in
the region of overlap. Section 3 describes the group selection 
algorithm.
Section 4 discusses the homogeneity of the group catalog across the two
underlying redshift surveys. We also examine the physical properties of 
the
optically selected groups and compare them with the x-ray emitting 
subset. In
Section 5 we conclude by comparing the USGC with similarly
constructed catalogs based on distinct redshift surveys including the 
ESO Key
Programme \citep{Ram99} and the LCRS \citep{Tuc00}.

\section{The Data}

We derive the group catalog from the official distributions of the UZC
(\url{http://cfa-www.harvard.edu/~falco/UZC}/) and SSRS2 catalogs.
We retrieve the SSRS2 from CDS using VizieR ~\citep{Och00}.
In the UZC catalog \citep{Fal99} we use only the $>$98\%  complete 
region limited by
$-2.5^{\circ} \leq  \delta_{1950} \leq 50^{\circ}$ and $8^{h} \leq
\alpha_{1950} \leq 17^{h}$ in the North Galactic Cap and  by $20^{h} 
\leq
\alpha_{1950} \leq 4^{h}$ in the South Galactic Cap.  We  discard the 
regions
$-13^{ \circ} \leq b \leq 13^{\circ}$ and $\alpha_{1950} \geq 3^h$ in 
the
southern galactic region because of the greater galactic absorption 
there
\citep{PTH01}.  We consider only galaxies with $cz < 15000$ km
s$^{-1}$.  There are 13333 galaxies in the subset of the UZC catalog we 
analyze.

The boundaries of the SSRS2 \citep{daC98} are $-40^\circ \leq \delta
\leq -2.5^\circ$ and $b_{II} \leq 40^\circ$ for SSRS2 South and $\delta 
\leq
0^\circ$ and $b_{II} \geq 35^\circ$ for SSRS2 North. Here, too, we 
consider only
galaxies with $cz < 15000$ km s$^{-1}$. 
The UZC and SSRS2 catalogs have slightly different magnitude systems, 
$m_{Zw}$
and $m_{SSRS2}$ respectively.  Detailed discussions comparing  $m_{Zw}$ 
and
$m_{SSRS2}$ are in ~\citet{Alo94} and, more recently, in 
~\citet{Mar98}. We use
magnitudes as listed within the UZC and SSRS2 catalogs and apply the
appropriate luminosity functions.

The SSRS2 and UZC overlap in the declination range $-3.8^\circ < 
\delta_{1950}
\leq 0^\circ $  and $140^\circ < \alpha_{1950} < 240^\circ$.  In this 
region the
UZC contains 472 galaxies, about twice as many galaxies as SSRS2 (260
galaxies).  Out of the 472 UZC galaxies in the Overlap Region (OR),
223 appear in both catalogs (we call this set of galaxies \uboth; and 
the remaining
249 galaxies UZC galaxies make up the sample \uonly).
We use the DSS to verify that the UZC objects omitted
from the SSRS2 are  galaxies. We note that 37 SSRS2 galaxies have no
UZC counterpart. Of these 37 galaxies, 10 are very bright NGC galaxies 
also within UZC, but outside the half arcminute radius we search for 
UZC counterparts. These 10 galaxies are so bright and rare that they do not enter
the magnitude comparison below. Extending our search to a 6 arcmin radius 
does not yield counterparts for the remaining 27 galaxies. These 27 galaxies
have apparent magnitudes spread throughout the SSRS2 range; the sample of
these galaxies is too small to draw any statistically interesting
conclusions and they have little effect on the analysis below.

The issue of completeness to the magnitude limit is of obvious 
importance for
the normalization of the luminosity function and potentially for the
determination of  large-scale structure parameters.
Several authors have addressed the problem of the ``quality''
of \mzw \citep{Huc76, Bot90, Tak95, Gro99, Gat00} and of \mss
\citep{Alo93, Alo94}.  Most of these authors
use CCD photometry to calibrate either m$_{Zw}$ or m$_{SSRS2}$. The
task is difficult because the CCD photometric samples are limited and 
there are  biases that can be introduced by
the selection of the calibrating galaxies and by the choice of 
photometric
techniques. It is perhaps not surprising that results are mostly 
discordant.

Given the lack of a ``final word'' on the two magnitude systems, it
is worth investigating the difference between galaxy counts in the OR.
For the 223 galaxies in the \uboth sample, we can compare  \mss to \mzw 
directly;  this sample is similar in size
to those in previous studies. Of course,  extension of the conclusions 
throughout the catalogs may not be
warranted.

The 223 \uboth galaxies are brighter, on
average, than the \uonly galaxies: only 25 \uboth galaxies
are fainter than \mzw =
15.20;  170 \uonly galaxies are fainter than this limit. The 
distribution of \mzw in the OR is very similar to the
distribution of \mss in the same region.
The only significant difference occurs
at the magnitudes fainter than \mzw = 15.2, more frequent among Zwicky magnitudes. 

We suggest that \mss and \mzw are not
equivalent and that SSRS2 magnitudes are systematically brighter than 
UZC magnitudes, at least in the OR.  A least
squares fit to the relationship \mzw vs. \mss
for \uboth galaxies yields
\mzw = 0.78 \mss + 3.16. We use this relation to transform \mss into new
magnitudes, \msst, that can be  compared directly to \mzw.
The new magnitude
limit is \msst = 15.25. This brighter limit is the
true limit that should be used for \mzw when comparing SSRS2 and UZC
magnitudes.

Remarkably, the magnitude limit \mzw $\leq 15.25$ leaves 277 UZC 
galaxies
in the OR,  very close count the 260 SSRS2 galaxies. Furthermore, a KS 
test
shows that the distributions of \mzw and \msst have a  probability 
greater than
98\% of having been drawn from the same parent population. 

The results we obtain in the OR may be valid everywhere else in the two
surveys. In fact, the distributions of \mss and \mzw within the OR do 
not
differ significantly from the magnitude distributions of the entire 
SSRS2 and
UZC, respectively. If we set \mzw = 15.25 as the new magnitude limit for
the whole UZC, there are 4842 galaxies in the UZC; there are 4824  
SSRS2 galaxies. The solid angle of both regions of the sky is about
1.6 sr. Thus the angular number densities of galaxies in
the two surveys {\bf coincide}. With
the original fainter UZC limiting magnitude, the angular
densities of the UZC and SSRS2
are 5200 galaxies sr$^{-1}$ and 3090 galaxies sr$^{-1}$ respectively.

Our analysis of the OR indicates that the UZC is a deeper survey than 
the
SSRS2 and that a linear transformation is needed to make \mzw and \mss
comparable. We note that \citet{Alo94} also analyze galaxies in common 
between
Zwicky's CGCG \citep{Zwi} and SSRS2. However, these authors focus on
the \uboth sample and ignore the presence of the large sample of \uonly
galaxies.
Without taking the \uonly galaxies into account, and ignoring possible 
scale errors,
they conclude that the average difference between \mss and \mzw is of 
the order of 0.1 mag and that it can be ignored in the analyses of UZC 
and SSRS2.   
Because we cannot verify that the results of our analysis of the OR
can be extended robustly over the entire UZC or SSRS2, we
do not attempt to unify selection
functions but rather use the appropriate selection function in each 
survey.
Relative to their selection functions, groups in the two surveys have 
similar properties.  Indeed, we will show below that the number of 
groups {\it relative to the underlying
galaxy distribution} and their main physical parameters are not 
significantly
different within the UZC and SSRS2 surveys.  
If the transformation between \mzw and \mss
is correct throughout the catalog, some ``absolute'' properties of 
groups will differ somewhat in the approach we
take here.
For example,  the average galaxy density within individual groups will 
be slightly different in
the UZC and SSRS2. We note, however, that
other uncertainties than magnitude scale
errors may dominate the identification of physically bound systems of 
galaxies
and the determination of their masses. We do not know either the
``true'' distances of galaxies independent of their velocities and thus
we do not have an estimate of the  ``true''
spatial galaxy density.  
It is beyond the scope of this paper to determine the ultimate 
underlying
reason for the discrepancy between the two catalogs in the overlap 
region. It
is also difficult to evaluate the propagation of these differences.
Progress will be made with large digital sky surveys, like 2MASS
and the on-going SDSS,
which will provide homogeneous and uniform CCD photometry and galaxy 
catalogs
over large regions of the sky.  Here we discuss groups  that have been 
used in a number of  different analyses, we take the magnitudes in the
two catalogs at face value and make no transformation. 
We use the UZC catalog in the
overlap region. The difference  in selection of galaxies within UZC and 
SSRS2
in the overlap region confuses the definition of groups at the border 
between
the two surveys; to avoid this problem we introduce a half degree gap 
between
the two surveys. In conclusion, the SSRS2 catalog we analyze consists 
of 4824
galaxies.

Figure 1a shows the distribution on the sky of  the  UZC and SSRS2 
galaxies
included in our analysis. We also plot (large dots) the position of the
Abell/ACO clusters within the volume of the survey \citep{And91}.  We 
download
Andernach's updated electronic table from CDS using VizieR 
~\citep{Och00}.
Several Abell/ACO clusters lie just outside the border of the SSRS2 
survey.  We
discuss the consequences of omission of these dense regions below.

\section{The Algorithm}

Friends-of-friends algorithms (FOFA) are now a standard approach to 
identifying
systems of galaxies in a redshift survey. We apply the FOFA originally 
proposed
by \citet{HG82} as implemented by \citet{Ram97}. 
The ``linking'' parameters $D_L$ and $V_L$ characterize the
FOFA search:  for each galaxy in the catalog, the FOFA
identifies all other galaxies with a projected separation
$$D_{12} \leq (V_1+V_2)\sin(\theta/2)/H_o\le D_L(V_1,V_2,m_1,m_2)
\eqno (1)$$
and a line-of-sight velocity difference

$$V_{12} \leq |V_1-V_2|\le V_L(V_1,V_2, m_1, m_2). \eqno (2)$$

Here, $V_1$ and $V_2$ are the velocities of the two galaxies in the 
pair,
$m_1$ and $m_2$ are their magnitudes,
$\theta$ is their angular separation, and $H_o$ is the
Hubble constant.
We assume a Hubble constant
$H_\circ = 100 h$ km s$^{-1}$ Mpc$^{-1}$ with $h = 1$ when an explicit 
value
is required. All pairs linked by a common galaxy form a ``group''.

We estimate the limiting  density contrast as

$${\delta\rho\over\rho}={3\over 4\pi D_o^3}
{\left[\int_{-\infty}^{M_{lim}}\Phi(M) dM\right]^{-1}} - 1.
\eqno (3)$$

Here $\Phi(M)$ is the luminosity function for the sample,
$M_{lim}$ is the faintest absolute magnitude at which galaxies
in a sample with magnitude limit $m_{lim}$ are visible at a
fiducial velocity $V_F$ = 1000 \kms, and $D_o$ is the
linking parameter $D_L$ at $V_F$ ~\citep{Ram89}. 
We scale
scale $D_L$ and $V_L$  to keep the number
density enhancement, $\delta\rho/\rho$, constant.
The scaling is

$$D_L=D_o R \eqno (4)$$

and

$$V_L = V_o R, \eqno (5)$$

where

$$R=\left[\int_{-\infty}^{M_{lim}} \Phi(M)
{dM}/\int_{-\infty}^{M_{12}} \Phi(M) {dM}\right]^{1/3} \eqno (6)$$

and

$$M_{12}= m_{lim}-25-5 \log((V_1+V_2)/2H_o) \eqno (7) $$

The ratio $D_L/V_L$ is  constant.

We define an association of three or more galaxies as a
``group''. We  consider only groups with mean velocities in the range 
500 \kms
$< V < 12000$ \kms. This value allows a straightforward comparison with 
the majority of published results on subsamples of CfA groups.  
For the UZC, the \citet{Sch76} form of the  galaxy luminosity function 
(LF)
has $M_{\star}=-19.1$, $\alpha=-1.1$, and $\phi_{\star}=0.04\ h^3$ 
Mpc$^{-3}$.
We obtain these values of the parameters by convolving an 0.3 magnitude
Gaussian uncertainty with the LF determined by \citet{Mar94} for a very
similar sample.  There is no significant difference between the {\it 
old}
\citet{Ram97} groups the {\it new} UZC groups where these
two samples overlap. Any differences result from the use of the more
accurate UZC coordinates and redshifts.  
For the SSRS2, the luminosity function parameters are 
$M_{\star}=-19.73$,
$\alpha=-1.2$, and $\phi_{\star}=0.013\; h^3$ Mpc$^{-3} $. We obtain 
these parameters by convolving the 0.3 magnitude uncertainty with the LF 
determined
by \citet{Mar94}.

We  set $D_0$ to correspond to a density contrast threshold $\delta 
\rho /
\rho$ = 80 within both the UZC and SSRS2.  Because the luminosity 
functions differ slightly
for the two catalogs, the linking parameters differ somewhat at the 
reference
velocity, $V_F$.  For both surveys, the fiducial linking parameters are 
approximately
$D_{0} = 0.25 h^{-1}$ Mpc  and $V_{0}=350$ \kms at $V_F = 1000 $ \kms. 
\citet{Ram97} discuss how group catalogs vary
with $\delta \rho /\rho$ and $V_{0}$ . They conclude that group 
properties are statistically stable for $\delta \rho /\rho$ in the
range 80 to 160. The choice $\delta \rho /\rho$ = 80 guarantees 
inclusion of the loosest systems and minimzes the chance splitting of 
the richest systems. The choice of $V_{0}$ is more critical. 
\citet{Ram97}
adopt the  value we use here based on the following considerations:
a) for this choice of  $V_{0}$, the velocity dispersions of the richest 
systems identified with FOFA
match the dispersions obtained from much larger samples (e.g. 
\citet{Zab93}),
b) observations show that groups obtained with $V_{0}=350$ \kms are
stable against inclusion of fainter members \citep{Ram95, Ram96},
c) analysis of the performances of FOFA on cosmological n-body 
simulations
show that $V_{0}=150$ \kms biases velocity dispersions toward low values
and that $V_{0}=550$ \kms produces highly inaccurate groups 
\citep{Fr95a}.

In a group catalog selected from a redshift survey with a FOFA, some 
fraction
of the groups are accidental superpositions.  We have two measures of 
the
fraction of true physical systems in the USGC catalog. \citet{Ram97} 
use geometric
simulations of the large-scale structure in the Northern UZC region to
demonstrate that at least  80\% of the groups with five or more members 
are probably
physical systems. ~\citet{Dia99} find similar results in their analysis
of mock CfA surveys extracted from n-body simulations. \citet{Mah00} 
cross-correlate a large fraction of
these richer groups with the ROSAT All-Sky Survey (RASS). A total of
61 groups, the
``RASSCALS'', in the \citet{Mah00} sample have associated extended
x-ray emi+ssion. The presence of hot x-ray emitting gas bound in the 
group
potential well is a confirmation of the physical reality of the system.
\citet{Mah00} use the groups detected in the x-ray to show that a
minimum fraction of 40\% of the groups in the subsample are similar 
x-ray
emitting systems undetectable with the RASS; thus they set a lower 
limit of
40\% on the fraction of real physical systems in the group catalog. We 
conclude
that 40-80\% of the groups with five or more members are real and it is
thus reasonable
to use their properties to derive physical constraints on the nature of 
groups
of galaxies. Of course, the statistical reliability is substantially 
less for
the 3 and 4 member groups. We include these systems as a finding list.

\section{The Group Catalog}

The USGC catalog contains 1168 groups, with the UZC and SSRS2 regions
containing 864 and 304 groups with a total of 5242 and 1604 member 
galaxies,
respectively. For each group, Table~\ref{tabgr} lists the group ID 
number, the
number of members within the survey, $N_{mem}$, the mean coordinates 
(J2000),
the mean heliocentric radial velocity, an unbiased estimate of the 
velocity
dispersion ~\citep{Led84} corrected for measurement errors and reduced 
to the
source redshift ~\citep{Dan80}, the virial radius, the virial mass, the
logarithm of the mass-to-light ratio,  the logarithm of the observed 
luminosity
and the logarithm of the total luminosity corrected for the unseen 
luminosity
to M$_Zw$ = -13.0 assuming a simple extrapolation of the relevant 
luminosity
function. We note that Table~\ref{tabgr}  lists the virial radius 
$R_{vir}=2 R_h N_{mem}/ (N_{mem}-1)$ \citep[see e.g.][]{Jac75, Roo78}; 
\citet{Ram97}
actually list $R_h$, the harmonic radius. Only the beginning of
Table~\ref{tabgr} is included here; the entire table is available
electronically.

Table~\ref{tabmem} is a sample of the list of group members; again the  
full list is
available electronically.  For each member we list the ID number of the 
parent
group, the galaxy coordinates (J2000) and magnitude as in the UZC or 
SSRS2, the
heliocentric velocity, and the velocity error.

We cross-identify groups in the USGC catalog  with the $N_{mem} \geq 5$ 
member
groups in \citet{Ram97} and with the x-ray detected RASSCALS.  Small 
changes in the
original galaxy catalogs and improved luminosity functions have small
statistical effects on the membership and/or physical parameters of the 
groups.
Table~\ref{xidrpg} lists the cross-identifications with \citet{Ram97}; 
Table~\ref{xidrass} contains
cross-identifications with the RASSCALS. All of the RASSCALS
detected in the x-ray are also in the USGC.  

\section{Properties of Groups}

Because of its size, the USGC catalog offers an opportunity to study a 
broad
range of physical properties of groups as a class of systems. We 
consider the
velocity dispersion, mass, and mass-to-light ratio here. \citet{Gir00}
evaluate the group correlation function.  \citet{Mah00} consider the
x-ray properties of groups in this sample. Pisani et al. (2002, in 
preparation) evaluate the multiplicity function.

Although the combined catalog covers 4.69 steradians, the sampling of 
the
distribution of rich clusters of galaxies is limited to a few systems.  
For the
SSRS2 which covers 1.56 steradians, the problem of the  small number of 
systems
with large velocity dispersion is most serious.  In their study of the 
pairwise
velocity dispersion for the CfA (a subset of the UZC) and SSRS2 South 
surveys,
\citet{Mar95} show that the dominant source of variance in $\sigma_{12}$
from one subsample to another is shot noise contributed by dense 
virialized
systems with large velocity dispersion.  We thus expect to uncover 
similar
limitations in our examination of the properties of groups. 
In Section~\ref{num_gr} we use the entire USGC catalog with three or 
more members to show
that selection of groups is homogeneous across the two catalogs. In 
section~\ref{phys_prop}
we specialize to the higher confidence groups with more than five 
members to
discuss the typical physical properties of the systems.  In 
section~\ref{rass} we show
that the  groups identified as extended x-ray sources have similar 
properties
in both the UZC and in the SSRS2. We conclude that, in spite of the 
differences
between the underlying galaxy catalogs, the group catalog is reasonably
homogeneous across the two surveys and provides a useful foundation for 
the
study of properties of these systems at low redshift.   

\subsection{The Number of Group Members}\label{num_gr}

Table~\ref{counts} lists the number of groups, members, non-members
($cz \leq 12000$ \kms) and galaxies within
the USGC. We also list these quantities separately for the two
surveys. From Table~\ref{counts}, we conclude that the UZC projected 
angular density of
groups, $\Sigma_{UZC} = 276$ sr$^{-1}$, is 1.4 times larger than the 
SSRS2
projected angular density, $\Sigma_{SSRS2} = 195$ sr$^{-1}$.  The 
Poisson
uncertainty on the angular densities is of the order of 10 groups 
sr$^{-1}$;
thus the difference between  $\Sigma_{UZC}$ and  $\Sigma_{SSRS2}$ is
significant.  In principle, the group-group correlation function 
increases the
uncertainty relative to the Poisson  estimate.  However, the amplitude 
of
the angular group-group correlation function, $s_0 = 8 h^{-1}$  Mpc 
\citep{Gir00}, is small compared with the size of the two surveys. 
Therefore we
expect that the effect of the   the group-group correlation function  
on the
uncertainty is negligible.

Our discussion of the UZC and SSRSS2 magnitude scales in Section 2, 
indicates
that the UZC is deeper than the SSRS2 and that a transformation
between the two magnitude systems may be required. In the spirit of 
that discussion, we ask what happens to the
relative surface number densities of groups
if we simply drop UZC galaxies
fainter than \mzw = 15.25 and construct a catalog with \nmem $\geq$
3. The group angular densities come into much better agreement, with
160 and 190 groups sr$^{-1}$ within UZC and SSRS2 respectively Shifting 
the limiting magnitude from \mzw = 15.25 to \mzw = 15.30 yields exactly 
the same angular densities. Although this apparently improved
agreement is enticing, we have no clear justification for extrapoolating
the results from the overlap region to the entire survey. 
The definition of survey boundaries may introduce bias in the survey. 
Several ACO clusters are just outside the boundary of SSRS2 (see Figure
1a). Because the correlation length of the richest systems is much 
larger
than for groups \citep[e.g.][]{Bah92,Bor99}, we consider the
projected angular density of poor systems as
a standard for comparing the two catalogs (without any correction for
possible differences in the magnitude systems) .  To estimate a 
``richness'', we set
all groups at the same fiducial velocity $V_F$ = 1000 km s$^{-1}$. We 
then use
the relevant luminosity function normalized to the  observed \nmem to 
compute
the expected number of members with $M_{Zw} \leq -14.5$, the absolute 
magnitude
corresponding to $m_{lim}=15.5$ at $V_F$. We thus obtain a corrected 
number of
members for each group, $N_{mem,V_F}$. 
The upper quartile of $N_{mem,V_F}$ is $N_{UQ} = 110$ members.  The 
projected
angular densities of groups poorer than $N_{UQ}$ are $\Sigma_{poor,UZC} 
= 188~
{\rm sr^{-1}}$ and $\Sigma_{poor,SSRS2} = 173 ~{\rm sr^{-1}}$, in 
reasonable agreement even
under the conservative hypothesis of Poisson uncertainties. The result 
is
insensitive to the exact choice of $N_{UQ}$.  This result suggests 
that, as in
the analysis of \citet{Mar95}, the richest systems account for the
different group abundances within the UZC and SSRS2.

To examine the homogeneity of the entire groups catalog further, we 
examine the
ratio between the number of groups, $N_{gr}$, and the number of 
galaxies,
$N_{gal}$.  In the redshift range $0$  km s$^{-1} < cz < 12000$ km 
s$^{-1}$,
there are 4484 galaxies within the SSRS2 and 12186 galaxies within the 
UZC
yielding $N_{gr} / N_{gal}$ = 0.068 and 0.071 respectively, with a 
Poisson
uncertainty of order $\pm$ 0.004. Figure 2 shows that these ratios are
approximately constant for the two surveys over a wide range  of 
redshifts. At
radial velocities $\gtrsim 10000$ km s$^{-1}$, groups become relatively 
less
abundant within the SSRS2, but the significance of the difference is 
low. The
$\pm 1 \sigma$ Poisson error bars in  Figure 2  are underestimates 
because
groups and galaxies are clustered.  Figure 2 supports the hypothesis
that $\Sigma_{UZC}$ and $\Sigma_{SSRS2}$ differ only because of ``fair
sampling'' issues: relative to the number of galaxies in each survey, 
the
abundance of groups is essentially the same.

The number of members relative to the galaxy population in the two 
surveys
gives another measure of the similarity of the group catalogs.  In 
principle,
even if our test of the number of groups is satisfactory, the 
distribution of
``richness'' could differ in the two surveys.

To explore this issue, we consider the fraction of galaxies in groups. 
There is a total of 5242 member galaxies in the UZC groups, corresponding to a
fraction $f_{UZC} = \Sigma_i N_{mem}^i/N_{gal }= 0.430 \pm 0.005$ of the total
number of galaxies within $0$ km s $^{-1} < cz < 12000$ km s$^{-1}$.  In the
SSRS2, $\Sigma_i N_{mem}^i = 1604$ and $f_{SSRS2} = 0.358 \pm 0.009$.  Taken at
face value, these fractions imply a difference between  the SSRS2 and the UZC.
From previous analyses of these catalogs we suspect that the source of the
difference here is shot noise in the abundances of the richest systems in the
two underlying redshift surveys. To investigate this possibility, Figure 3
shows the fraction of galaxies in groups, $f(cz)$, in redshift bins (we count
all members of a group within the redshift bin containing the group mean
velocity): from 2000 km s$^{-1}$ to 12000 km s$^{-1}$ $f_{UZC}(cz)$ (thick
solid line) is increasingly larger than $f_{SSRS2}(cz)$ (thin solid line).
Because the width of the redshift bins is relatively close to the correlation
length of groups, we do not use Poisson error estimates.  We estimate the
1-$\sigma$ error-bars in Figure 3 by running a  bootstrap procedure 10000 times
in each $cz$ bin. For each bootstrapped sample we compute the total number of
members and compute $f^i(cz), i=1,..,10000$.  Error-bars in Figure 3 are
1-$\sigma$ of the bootstrapped distribution of $f(cz)$.  To see whether the
difference in $f(cz)$ for all groups is attributable to the richest systems,
Figure 3 also shows $f(cz)$ for the poor groups with $N_{mem,V_F}^i < 110$.
Without the rich systems, $f_{UZC}(cz)$, filled circles, and $f_{SSRS2}(cz)$,
empty circles, are in reasonable agreement.

We conclude that groups within the SSRS2 and UZC are very similar for 
both
their abundance and membership  {\it relative to the galaxy 
distribution}.  If
we exclude the richest systems, the fraction of galaxies identified as 
group
members is the same at any redshift in both surveys.  This result 
indicates
that observational biases in the UZC relative to the SSRS2 do influence 
the
identification of groups significantly.  
\subsection{The physical properties of groups}\label{phys_prop}

To examine some of the physical properties of groups, we restrict the 
analysis
to higher confidence systems with at least five observed members.  
There are
313  such groups in the UZC and 98 in the SSRS2. Table~\ref{grprop} 
lists the
median and 90\% confidence levels of the velocity dispersion, 
$\sigma_{cz}$,
the virial radius, $R_{vir}$, the mass, $M_{vir}$, and the 
mass-to-light ratio,
$M/L$. We compute the luminosity from the original Zwicky magnitudes 
for the
UZC; we use the SSRS2 catalog magnitudes for SSRS2 groups. In combining 
the
catalogs we ignore possible differences between the two magnitude 
systems.  We
list the median properties for the combined group sample and for the 
individual
UZC and SSRS2 samples.

Table~\ref{grprop} shows that, in spite of differences in the 
construction of
the UZC and SSRS2, the median properties of groups are the same.  We 
use KS
tests to compare the distributions of these properties of groups within 
the UZC
and SSRS2. We find that, nothwhistanding the almost coincident medians, 
the distributions of
$\sigma_{cz}$ and $M$ within UZC and SSRS2 are unlikely to be drawn 
from the
same parent distribution (at the  $>$99\% confidence level).

To assess the source of the results for the $\sigma_{cz}$ and $M$
distributions, Figure 4 shows the integral distributions of  
$\sigma_{cz}$ for
systems within the UZC (thick histogram) and the SSRS2 (thin 
histogram).  Again
we see the issue we have uncovered before: groups with high velocity 
dispersion
explain the difference between the two distributions.

To explore this issue further we divide the group catalog into rich and 
poor
subsamples. Again, we divide the sample based on the third quartile of 
the
corrected richness appropriate for groups with five or more members, 
$N_{UQ,5}
= 200$ (see section~\ref{num_gr}), we drop eight groups with $cz \leq 
1000$
\kms (these systems are within the Local Supercluster).  Figure 4 
compares the
integrated distributions of $\sigma_{cz}$ for the poor subsamples of 
groups. A
KS test shows that these subsamples are unlikely to be drawn from 
different
distributions (P(KS) = 39\%).  The situation is similar for the 
distribution of
$M$. 
Table~\ref{grprop} explores the variation in median physical parameters 
of the
four poor/rich group subsamples (excepting the  small subsample of five 
SSRS2
rich groups). The median quantities for the poor subsamples are in 
excellent
agreement with the full SSRS2 sample but differ from  the total UZC. As
expected, the rich UZC groups have a significantly larger median 
$\sigma_{cz}$
and $M$. In contrast the median $M/L$ is remarkably consistent for all
subsamples.

We conclude that both the UZC and SSRS2 provide fair samples of poor 
systems.
These systems have similar statistical properties in the two catalogs. 

\section{X-ray groups}\label{rass}

A subsample of 61 groups in the USGC are extended x-ray sources 
detected in the
RASS \citep{Mah00}. These groups are an important subset because the 
x-ray
emission from gas presumably held in the potential well of the group 
guarantees
that these groups are real physical systems. It is therefore 
interesting to
compare  their physical parameters to those for all groups in the 
catalog. 
Of the 61 groups, 16 are within SSRS2 and 45 within UZC. Therefore 
about $\sim$
15\% of the groups in both samples are detected in the RASS.  Among the 
x-ray
emitting groups, 41 are poor according to the definition in
section~\ref{phys_prop}. 
Table~\ref{grprop} lists the physical parameters of the entire 
x-ray-emitting
subsample.  It is not surprising that median $\sigma_{cz}$ is 
intermediate
between the rich and poor subsamples. The x-ray detected groups are
systematically more massive than the typical system in the poor groups 
catalog.
One concern is that this difference is simply a selection effect; the 
x-ray
identification might pick out more distant, richer groups. However, a 
KS test
demonstrates that the velocity distribution of the poor groups detected 
in the
x-ray poor groups is consistent with the entire poor USGC sample. This
consistency is reassuring; it indicates that the sample of poor groups 
with
five or more members provides a foundation for the assessment of  the 
physical
properties of these systems and their member galaxies. 

\section{Conclusion}

The UZC and SSRS2 redshift surveys provide the basis for the USGC 
catalog which
covers 4.69 steradians to 12000 \kms. The FOFA identifies a homogeneous 
set of
411 systems with more than five observed members.  For completeness and 
for
verification of the uniformity of the catalog we include all 1168 
groups with
3 or more members.

For poor systems in the first three quartiles of the ``richness''
distribution, the surface number density of groups, the fraction of 
galaxies in
groups, and the properties of groups are essentially the same for both 
the UZC
and for the SSRS2 . In the upper quartile, the catalogs differ: only 
the UZC
contains a substantial number of these systems.  As in \citet{Mar95} we
attribute this difference to shot noise in the sampling of the most 
massive
systems in the smaller SSRS2 survey. Both the UZC and the SSRS2  appear 
to be
large enough to yield a fair sample of poor systems. 
The UZC and
SSRS2 catalogs overlap in a region which covers $\sim 0.11$ sterad.
In this region, the UZC contains almost twice as many
galaxies than SSRS2. We obtain a linear transformation between \mzw and 
\mss for galaxies that have both magnitudes.
The new
\msst is  directly comparable to \mzw, and its faintest
value is \msst = 15.25. Cutting the UZC at this shallower limit brings 
the UZC and SSRS2 catalogs  into remarkable agreement in the overlap 
region.

If we extend the limit \mzw = 15.25 to the whole UZC, the global
angular number densities of galaxies in the UZC and SSRS2 become 
comparable. The angular number density of groups in the
shallower UZC catalog is essentially the same as
for the SSRS2.
However, the main physical properties of groups do not change with 
re-identification within the shallower UZC or within SSRS2 using
\msst instead of \mss. In these modified
catalogs it remains true that the UZC has a larger
number of rich, high velocity systems than SSRS2 and that the
\sv distributions of ``poor'' groups in the two surveys
are likely to be drawn from the same parent distribution.

Extended x-ray emission associated with a subset of the USGC with more 
than
five observed members provided reassurance that a large fraction of the 
systems
we identify are true physical associations.  Statistical comparison of 
the
x-ray and optical selection suggests that at least 40\% of the groups we
identify have x-ray properties similar to the RASS sample. Geometric 
analyses
of a portion of the UZC suggest that $\sim 80\%$ of the groups are true
physical systems. Comparison with n-body simulations confirm this upper 
limit.

\citet{Ram99} used the same procedure we apply here to derive a group 
catalog
from a deeper redshift survey of an independent region, the ESO Slice 
Project
(ESP). The ESP group catalog includes a remarkably similar fraction of 
the
galaxies in the redshift survey. Furthermore, the median velocity 
dispersion of
groups with five or more members is 272 \kms with an interquartile 
range of
(178, 379) \kms essentially the same as the comparable USGC sample with 
a
median velocity dispersion of 264 \kms and an interquartile range of 
(150, 407)
\kms.  Other physical properties of the groups in this deeper catalog 
are also
similar to those for the USGC. \citet{Tuc00} compare the properties of 
groups
derived from the Las Campanas Redshift Survey \citep[][LCRS;]{Lin96} 
with a
broad range of catalogs including the subset of the USGC discussed by
\citet{Ram97}. \citet{Tuc00} conclude that the properties of their 
groups at
median redshift $z \sim 0.1$ are similar to the systems in the nearby 
samples. 
Groups have traditionally provided a basis for estimating the mean mass 
density
of the universe. Now there are better methods of approaching this issue
~\citep[e.g.][]{Rie98, Per99, Bal00, deB00, Mel00}.  However, the 
existence of
algorithms which produce uniform catalogs of systems, many of which are 
true
physical associations, is a basis for exploration of groups as 
environments for
galaxy evolution and for investigation of groups as tracers of 
large-scale
structure in the universe.  The USGC provides an extensive low redshift
baseline for these studies.  
\acknowledgments

We thank Antonaldo Diaferio, Scott Kenyon, Michael Kurtz, and Andisheh 
Mahdavi
for discussions and inspiration to complete the catalog. We thank the
ananymous referee for urging us to clarify a number of issues thus
strengthening the paper. We acknowledge
the Space Telescope Science Institute for the usage of the DSS database.
The Smithsonian Institution, the Italian MIUR, and the Regione F.V.G. 
supported this research.

\clearpage

\begin{figure}
\plotone{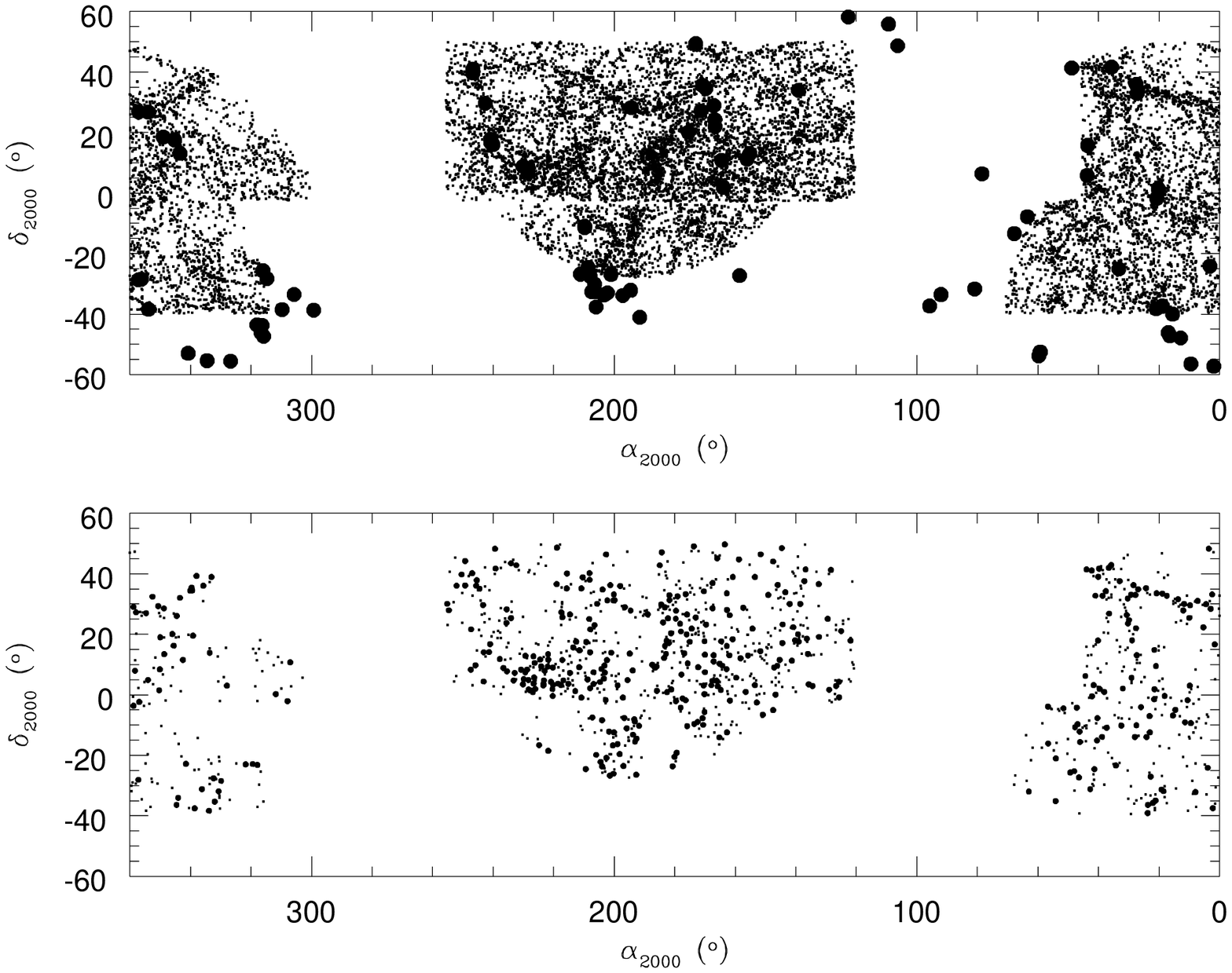}
\caption{(a) Right ascension and declination of UZC and SSRS2  
galaxies. Large dots are Abell/ACO
clusters with measured redshift cz $\le$ 12000 \kms. (b)  Right 
ascension and declination of the USGC groups. Smaller points
are triplets and quadruplets; larger points are groups with at least
five members in the redshift catalog.
\label{fig1}}
\end{figure}
\clearpage

\begin{figure}
\plotone{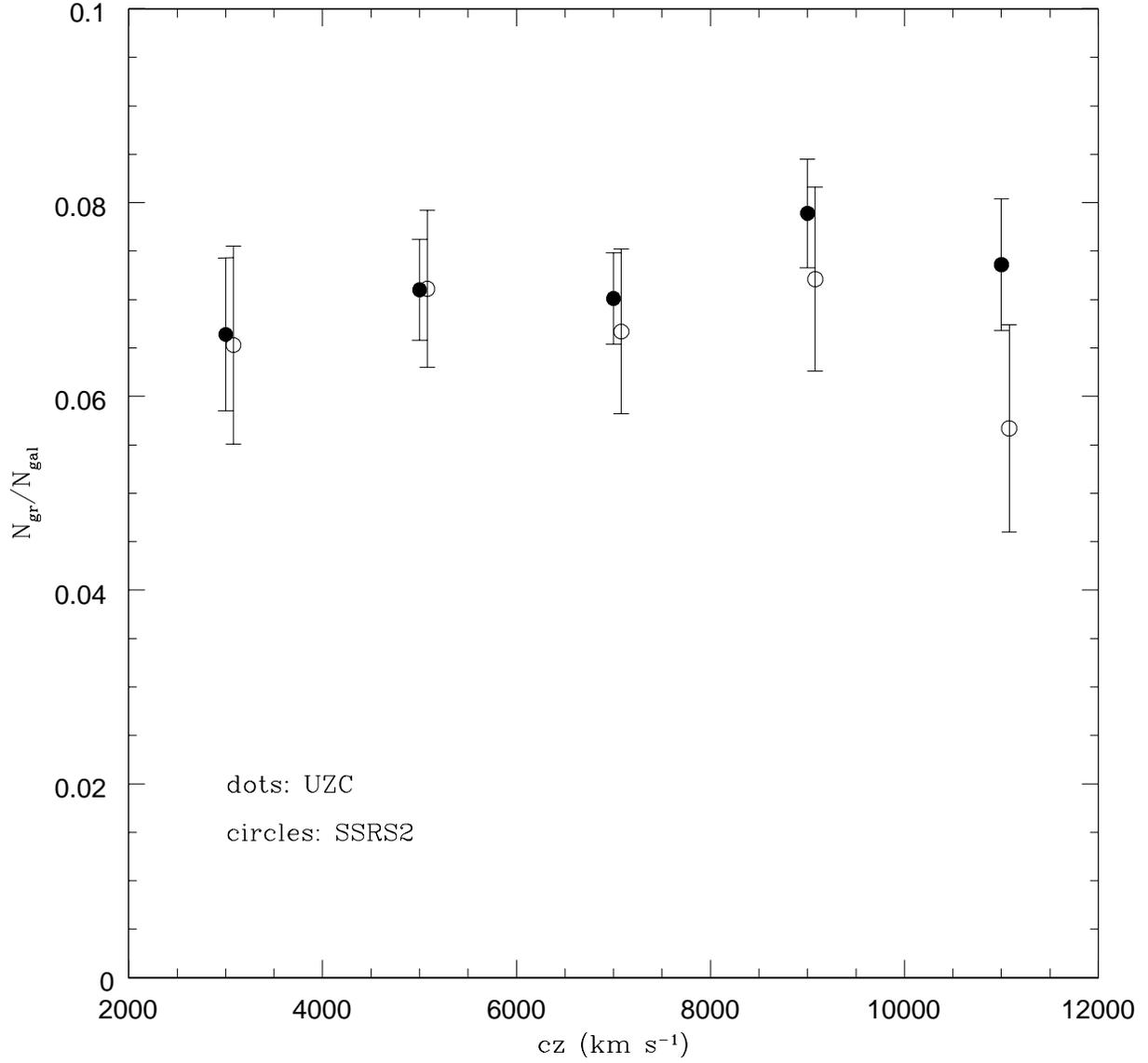}
\caption{ Ratio $N_{gr} / N_{gal}$ computed
in redshift bins. Error bars indicate $\pm 1 \sigma$,
Poisson errors. Filled symbols are for UZC,
empty symbols for SSRS2.
\label{fig2}}
\end{figure}
\clearpage

\begin{figure}
\plotone{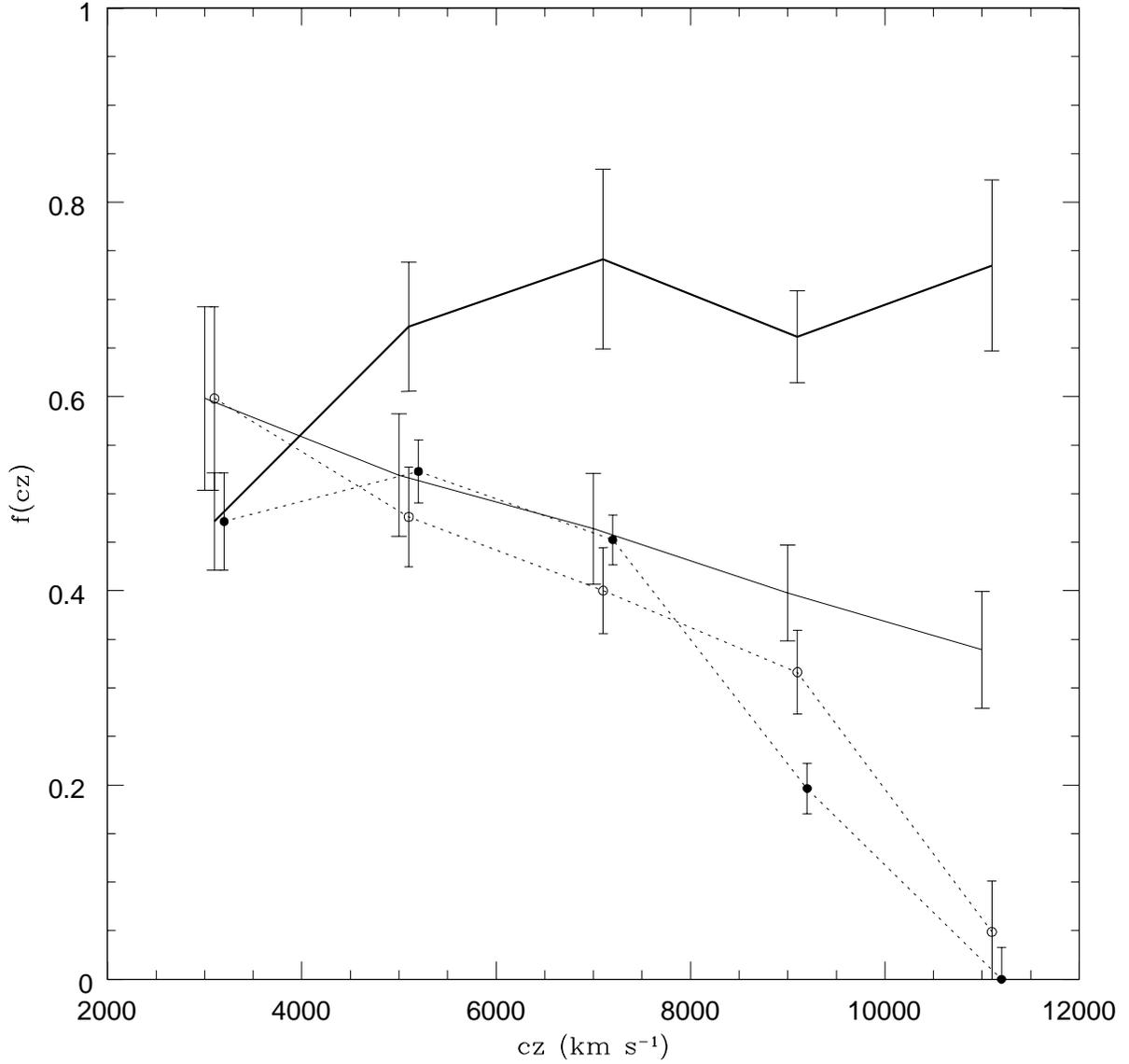}
\caption{Fraction of galaxies in groups, f(cz), within
UZC (thick solid line) and SSRS2 (thin solid line).
Error bars correspond to $\pm 1 \sigma$ estimated using 10000 bootstrap 
samples. Dotted lines represent f(cz) computed
for ``poor'' groups within
UZC (filled symbols) and SSRS2 (empty symbols) - error bars have
been omitted for clarity. \label{fig3}}
\end{figure}
\clearpage

\begin{figure}
\plotone{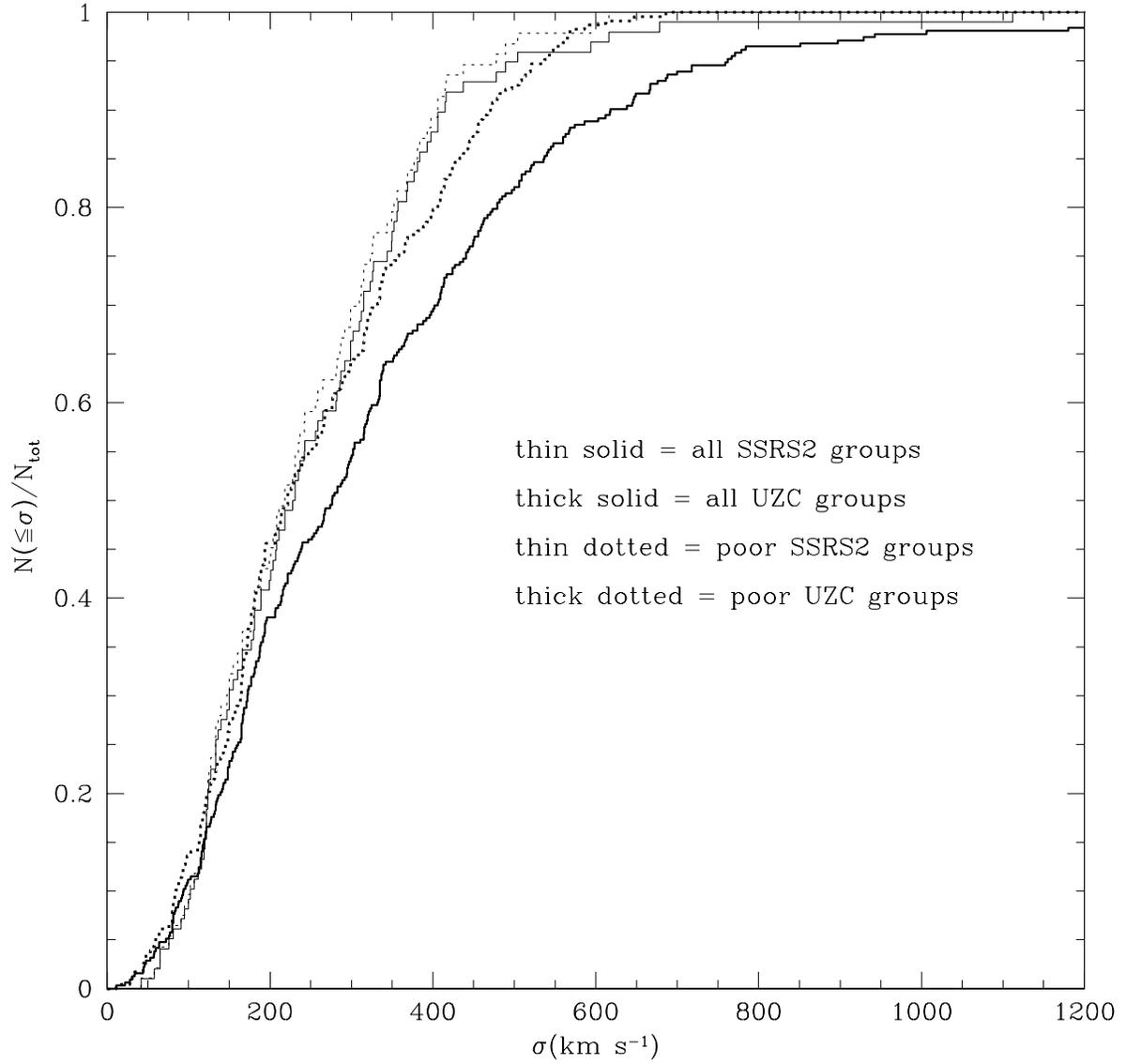}
\caption{Normalized integral
distributions of  $\sigma_{cz}$ for systems within the UZC
(thick histogram) and
SSRS2 (thin histogram). Solid lines are for all groups, dotted lines 
for ``poor'' groups.
\label{fig4}}
\end{figure}
\clearpage

\begin{deluxetable}{rcrrrrrrr}
\tablecolumns{9}
\tablewidth{0pc}
\tablecaption{ UZC-SSRS2 Group Catalog.\tablenotemark{a}
  \label{tabgr}}
\tablehead{
\colhead{ID} & \colhead{$N_{mem}$} & \colhead{$\alpha_{2000}$} &
  \colhead{$ \delta_{2000}$}
  & \colhead{$c\bar{z}$} & \colhead{$ \sigma_{cz}$} &    
\colhead{$R_{vir} $} & \colhead{$\log(M)$ }& \colhead{$\log(M/L)$}\\
\colhead{~~} &\colhead{~~}  & \colhead{$hh$~$mm$~$ss$}
  & \colhead{$^\circ$~$'$~$''$}
& \colhead{$km s^{-1}$} & \colhead{$km s^{-1} $} &    
\colhead{$Mpc~h^{-1}$} & \multicolumn{2}{c}{~~}
}
\startdata
  U001&   3&00~00~05.6& 26~11~25& 7450& 482& 0.61&13.99& 3.15 \\
  U002&   4&00~00~07.7& 32~45~58&10190& 202& 0.58&13.22& 1.73 \\
  U003&   3&00~00~43.0& 28~23~45& 8560& 391& 0.25&13.42& 2.31 \\
  U004&   3&00~01~41.8& 13~03~54& 5378& 142& 0.50&12.85& 2.23 \\
  U005&   4&00~05~38.3& 05~09~32& 5465& 265& 0.58&13.45& 3.02 \\
  U006&   5&00~06~07.7& 16~40~50&  972& 136& 1.07&13.14& 3.60 \\
  U007&   3&00~08~05.7& 47~07~42& 5173& 139& 0.58&12.89& 2.44 \\
\enddata
\tablenotetext{a}{The complete version of this table is in
the electronic edition of the Journal.  The printed edition contains 
only a sample.}
\end{deluxetable}
\clearpage
\begin{deluxetable}{rcrrrrr}
\tablecolumns{7}
\tablewidth{0pc}
\tablecaption{USGC Member Galaxies.\tablenotemark{a}\label{tabmem}}
\tablehead{
\colhead{ID} & \colhead{$N_{mem}$} & \colhead{$\alpha_{2000}$} &
  \colhead{$ \delta_{2000}$}
  & \colhead{$m$} &    \colhead{$cz$} & \colhead{$\epsilon_{cz}$}\\
\colhead{~~} &\colhead{~~}  & \colhead{($hh$~$mm$~$ss$)}
  & \colhead{($^\circ$~$'$~$''$)}
& \colhead{~~} & \colhead{($km s^{-1} $)} &    \colhead{($km s^{-1} $)}
}
\startdata
  U001&   3&00~00~31.7&+26~18~19& 14.7& 7754&  52 \\
  U001&   3&00~00~31.4&+26~19~30& 14.7& 7637&  14 \\
  U001&   3&23~59~13.7&+25~56~26& 15.5& 6959&  37 \\
  U002&   4&00~00~16.2&+32~47~33& 15.1&10332&  34 \\
  U002&   4&00~00~14.8&+32~49~55& 15.1& 9956&  69 \\
  U002&   4&00~00~09.2&+32~44~18& 14.9&10372&  19 \\
  U002&   4&23~59~50.6&+32~42~08& 15.3&10100&  52 \\
\enddata
\tablenotetext{a}{The complete version of this table is in
the electronic edition of the Journal.  The printed edition contains 
only a sample.}
\tablecomments{'U' members are UZC galaxies, 'S' members are SSRS2 
galaxies. Magnitude $m$ is $m_{Zw}$ for UZC galaxies and
$m_{SSRS2}$ for SSRS2 galaxies}
\end{deluxetable}
\clearpage
\begin{deluxetable}{cc}
\tablecolumns{2}
\tablewidth{0pc}
\tablecaption{Crossreferences for RPG97 Groups \tablenotemark{a} 
\label{xidrpg}}
\tablehead{
\colhead{RPG97} & \colhead{UZCGG} \\}
\startdata
  RPG002& U185 \\
  RPG004& U189 \\
  RPG007& U193 \\
  RPG008& U192 \\
  RPG015& U200 \\
  RPG016& U203 \\
\enddata
\tablenotetext{a}{The complete version of this table is in
the electronic edition of the Journal.  The printed edition contains 
only a sample.}
\tablecomments{Only RPG97 groups with \nmem $\geq 5$.}
\end{deluxetable}
\clearpage

\begin{deluxetable}{cc}
\tablecolumns{2}
\tablewidth{0pc}
\tablecaption{Crossreferences for RASSCALS 
\tablenotemark{a}.\label{xidrass}}
\tablehead{
\colhead{RASSCALS} & \colhead{USGC} \\}
\startdata
  NRGb004&  U189 \\
  NRGb032&  U223 \\
  NRGb045&  U238 \\
  NRGb078&  U288 \\
  NRGb128&  U381 \\
  NRGb151&  U412 \\
\enddata
\tablenotetext{a}{The complete version of this table is in
the electronic edition of the Journal.  The printed edition contains 
only a sample.}
\tablecomments{Only Madhavi et al. 2000 groups with extended x-ray 
emission}
\end{deluxetable}
\clearpage

\begin{deluxetable}{lcrrrr}
\tablecolumns{6}
\tablewidth{0pc}
\tablecaption{Galaxy Counts.\label{counts}}
\tablehead{
\colhead{Survey} & \colhead{$\Omega$} & \colhead{$N_{gr}$} & 
\colhead{$N_{mem}$}
& \colhead{$N_{non-mem}$\tablenotemark{a}} & \colhead{$N_{gal}$} \\
\colhead{~~}&\colhead{($sterad$)} & \colhead{~~}&\colhead{~~}& 
\colhead{~~} &
\colhead{~~}}
\startdata
U+S   &   4.69   &  1168   &  6846    &  9882       &   16728 \\
UZC   &   3.13   &   864   &  5242    &  7020       &   12262 \\
SSRS2 &   1.56   &   304   &  1604    &  2862       &    4466 \\
\enddata
\tablenotetext{a}{$cz \leq 12000$ \kms}
\end{deluxetable}
\clearpage

\begin{deluxetable}{lccccc}
\tablecolumns{6}
\tablewidth{0pc}
\tablecaption{ Properties  of Groups with $N_{mem} \ge 
5$.\label{grprop}}
\tablehead{
\colhead{Survey} & \colhead{$N$} & \colhead {$\sigma_{cz}$}  & 
\colhead{$R_{vir}$} & \colhead{$log(M/M_{\odot})$} & 
\colhead{$log((M/L)/(M_{\odot}/L_{\odot}))  $} \\
\colhead{~~} & \colhead{~~} & \colhead {($km s^{-1}$)}  & 
\colhead{($Mpc~ h^{-1}$)} & \colhead{~~} & \colhead{~~} }
\startdata
USGC   &  411    &   264 (229,292)&   1.06 (0.98,1.14)&   13.67 
(13.59,13.76)&   2.63 (2.58,2.68)\\
UZC   &  313    &   283 (247,315)&   1.09 (1.01,1.17)&   13.73 
(13.64,13.84)&   2.63 (2.57,2.67)\\
SSRS2 &    98   &   229 (199,259)&   0.99 (0.89,1.13)&   13.59 
(13.39,13.69)&   2.67 (2.56,2.76)\\
RASSCALS & 61 &   409 (352,437)&   1.10 (0.94,1.28)&   14.02 
(13.91,14.17)&   2.64 (2.61,2.68)\\
UZC Poor   & 228 &   222 (194,254)&   0.93 (0.88,1.01)&   13.49 
(13.42,13.62)&   2.62 (2.56,2.69)\\
SSRS2 Poor &  93  &   218 (189,242)&   0.95 (0.88,1.11)&   13.56 
(13.37,13.65)&   2.67 (2.56,2.76)\\
UZC Rich   &  77 &   506 (411,603)&   1.84 (1.73,1.98)&   14.43 
(14.31,14.57)&   2.63 (2.53,2.67)\\
\enddata
\end{deluxetable}

\begin{thebibliography}{ }
\bibitem[Alonso et al. (1993)]{Alo93} Alonso, M. V., da Costa, L. N.,
          Pellegrini, P. S., \& Kurtz, M. J. 1993, AJ 106, 676
\bibitem[Alonso et al. (1994)]{Alo94} Alonso, M. V., da Costa, L. N.,
          Latham, D. W., Pellegrini, P. S., \& Milone, A. E. 1994, AJ 
108, 1987
\bibitem[Andernach (1991)]{And91} Andernach H., 1991, ASP Conf. Ser. 
15, 279,           (eds. D.W.Latham and L.N.daCosta), CDS VII/165a 
\bibitem[Bahcall \& West (1992)]{Bah92} Bahcall, N. A., \& West, M. 
J.          1992, ApJ 392, 419 \bibitem[Balbi et al. (2000)]{Bal00} 
Balbi, A., et al. 2000, \apjl ~545, L1
\bibitem[de Bernardis et al. (2000)]{deB00} de Bernardis, P. et al. 
2000,
          \nat ~404, 955
\bibitem[Borgani et al. (1999)]{Bor99}Borgani, S., Plionis, 
M.,          \& Kolokotronis, V. 1999, MNRAS 305, 866 \bibitem[Bothun 
\& Cornell (1990)]{Bot90} Bothun, G. D., \& Cornell, M. E.
	 1990, AJ 99, 1004
\bibitem[da Costa et al. (1998)]{daC98} da Costa L. N. et al. 1998, AJ 
116, 1 \bibitem[Carlberg et al. (2001)]{Car01} Carlberg, R. G., Yee, H. 
K. C.,
          Morris, S. L., Lin, H., Hall, P. B., Patton, D. R.,          
Sawicki, M., \& Shepherd, C. W.  2001, ApJ 552, 427 \bibitem[Danese et 
al. (1980)]{Dan80} Danese L., De Zotti G., \& di Tullio G.          
1980, A\&A 49, 137 \bibitem[Diaferio et al. (1999)]{Dia99} Diaferio, 
A., Kauffmann, G.,          Colberg, J.M., \&  White, S.D.M. 1999, 
MNRAS 307, 537
\bibitem[Falco et al. (1999)]{Fal99} Falco, E. E. et al.  1999, PASP 
111, 438
\bibitem[Frederic (1995a)]{Fr95a} Frederic, J.J. 1995a, ApJS, 97, 259 
\bibitem[Frederic (1995b)]{Fr95b} Frederic, J.J. 1995b, ApJS, 97, 275 
\bibitem[Garcia (1993)]{Gar93} Garcia, A. M. 1993 A\&AS 100, 47
\bibitem[Gatza\~naga \& Dalton (2000)]{Gat00} Gatza\~naga, E., \& Dalton, 
G. B. 2000,
	 MNRAS 312, 417
\bibitem[Girardi et al. (2000)]{Gir00} Girardi, M., Boschin, W., \&
          da Costa, L. N. 2000, \aa ~353, 57 \bibitem[Giuricin et al. 
(2000)]{Giu00} Giuricin, G., Marinoni, C.,          Ceriani, L., \& 
Pisani, A. 2000 ApJ 543, 178 \bibitem[Gourgoulhon et al. (1992)]{Gou92} 
Gourgoulhon, E., Chamaraux,
          P.,\& Fouque, P. 1992, A\&A 255, 69 \bibitem[Grogin \& Geller 
(1999)]{Gro99} Grogin, N. A., \& Geller, M. J. 1999,
          AJ 118, 2561
\bibitem[Huchra (1976)]{Huc76} Huchra, J.P. 1976, AJ, 81, 952 
\bibitem[Huchra \& Geller (1982)]{HG82} Huchra, J.P., \& Geller, M. J.
          1982, ApJ, 257, 423 \bibitem[Jackson (1975)]{Jac75} Jackson, 
J. C. 1975, MNRAS 173, 41P
\bibitem[Ledermann (1984)]{Led84} Ledermann W. 1984,          Handbook 
of Applicable Mathematics, Vol. VI, part A (Wiley and Sons)
\bibitem[Lin et al. (1996)]{Lin96} Lin, H., Kirshner, R. P., Shectman, 
S. A.,
          Landy, S. D., Oemler, A., Tucker, D. L., Schechter, P. 
L.          1996, ApJ 464, 60 \bibitem[Mahdavi et al.(2000)]{Mah00} 
Mahdavi, A., B\"{o}hringer, H., Geller,
          M.J., \& Ramella M. 2000, ApJ 534, 114  \bibitem[Maia et al. 
(1989)]{Mai89} Maia, M.A.G., da Costa, L. N., Latham,
          D. W. 1989, ApJS 69,809 \bibitem[Marzke et al. (1994)]{Mar94} 
Marzke R.O., Huchra, J. P., \&
          Geller M.J. 1994, ApJ 428, 43 \bibitem[Marzke et al. 
(1995)]{Mar95} Marzke, R. O., Geller, M. J., da Costa,
           L. N., \& Huchra, J. P.  1995, AJ 110, 477 \bibitem[Marzke 
et al. (1998)]{Mar98} Marzke R.O., da Costa L.N., \&          Geller 
M.J. 1998, ApJ 503, 617 \bibitem[Materne (1978)]{Mat78} Materne, J. 
1978, A\&A 63, 401 \bibitem[Melchiorri et al. (2000)]{Mel00}Melchiorri, 
A. et al. 2000,          \apjl ~536, L63
\bibitem[Nolthenius \& White (1987)]{NW87} Nolthenius, R., \& 
White,          S. D. M. 1987, MNRAS 225, 505 \bibitem[Ochsenbein et 
al. (2000)]{Och00} Ochsenbein, F., Bauer, P., \&          Marcout, J. 
2000, A\&AS 143, 221
\bibitem[Padilla et al. (2001)]{Pad01} Padilla, N. D., Merch\'{a}n, M. 
E., Valotto,          C. A., Lambas, D. G., \& Maia, M. A. G. 2001, ApJ 
554, 873 \bibitem[Padmanabhan et al. (2001)]{PTH01}Padmanabhan, N., 
Tegmark, M.,
          Hamilton, A.J.S., 2001, ApJ, 550, 52 \bibitem[Perlmutter, S., 
et al. (1999)]{Per99} Perlmutter, S., et al.          1999, ApJ 517, 
565 \bibitem[Pisani et al. (1992)] {Pis92} Pisani, A., Giuricin, G., 
Mardirossian,          F., \& Mezzetti, M. 1992, ApJ 389, 68
\bibitem[Ramella et al. (1989)]{Ram89} Ramella, M., Geller, M. 
J.,\&          Huchra, J.P. 1989, ApJ 344,57
\bibitem[Ramella et al. (1995)]{Ram95} Ramella, M., Geller, M. 
J.,          Huchra, J.P., \& Thorstensen, J. R. 1995, AJ 109,1469  
\bibitem[Ramella et al. (1996)]{Ram96} Ramella, M., Focardi, P.,
          Geller, M. J. 1996, A\&A 312, 745
\bibitem[Ramella et al. (1997)]{Ram97}Ramella, M., Pisani, A.,
          Geller, M. J. 1997, AJ 113, 483 (RPG97) \bibitem[Ramella et 
al. (1999)]{Ram99} Ramella, M. et al. 1999, A\&A 342, 1 \bibitem[Riess 
et al. (1998)]{Rie98} Riess, A. G. et al. 1998, AJ 116, 1009
\bibitem[Rood \& Dickel (1978)]{Roo78} Rood, Dickel 1978, ApJ 224, 724 
\bibitem[Schechter (1976)]{Sch76} Schechter, P.L. 1976, AJ, 203, 297 
\bibitem[Takamiya et al. (1995)]{Tak95} Takamiya, M., Kron, R. G., \& 
Kron,
          G. E. 1995, AJ 110, 1083
\bibitem[Trasarti-Battistoni et al. (1997)]{Tra97}Trasarti-Battistoni, 
R.,          Invernizzi, G., \& Bonometto, S. 1997, ApJ 475, 1 
\bibitem[Trasarti-Battistoni (1998)]{Tra98}Trasarti-Battistoni, R. 1998,
          A\&AS 130, 341
\bibitem[Tucker et al. (2000)]{Tuc00} Tucker, D.L. et al. 2000, ApJS 
130, 237 \bibitem[Tully (1987)]{Tul87} Tully, B. 1987, ApJ 321, 280 
\bibitem[Turner \& Gott (1976)]{TG76} Turner, E.L., \& Gott, J. R. 1976,
          ApJS 32, 409 \bibitem[de Vaucouleurs (1974)]{deV74} de 
Vaucouleurs, G. 1974,
          Stars and Stellar Systems, Vol.9, ed.
          A. and M. Sandage, \& J. Kristian          (Chicago:Univ. 
Chicago Press), 557
\bibitem[Vennik (1984)]{Ven84} Vennik, J. 1984, Tartu Astr. Obs. Publ., 
73, 1 \bibitem[White et al. (1999)]{Whi99} White, R. A., Bliton, M., 
Bhavsar, S. P.,          Bornmann, P., Burns, J. O.,          Ledlow, 
M. J., \& Loken, C. 1999, AJ 118, 2014 \bibitem[Zabludoff et al. 
(1993)]{Zab93} Zabludoff A., Geller, M. J.,
          Huchra, J. P. \& Vogeley, M. S. 1993, AJ 106, 1273
\bibitem[Zwicky et al. (1961---1968)]{Zwi}Zwicky F., Herzog E., Wild 
P.,          Karpowicz M. \& Kowal C.
          1961---1968, Catalogue of Galaxies and of Clusters          
of Galaxies (Pasadena:
          California Institute of Technology)

\end{thebibliography}
\end{document}